\documentstyle[11pt,newpasp,twoside,epsfig]{article}
\newcommand{\tfe}{1E~1048.1--5937}
\newcommand{\tfn}{1E~2259.1+586}
\newcommand{\soe}{RXS~J170849.0--400910}
\newcommand{\oft}{4U~0142+61}
\newcommand{\efo}{1E~1841--045}

\index{anomalous X-ray pulsars}
\index{4U 0142+61}
\index{1E 2259+586}
\index{phase-coherent timing}	
\index{magnetars}
\index{Rossi X-ray Timing Explorer}

\def\edcomment#1{\iffalse\marginpar{\raggedright\sl#1\/}\else\relax\fi}
\reversemarginpar

\begin{document}
\title{Long-Term Monitoring of Anomalous X-ray Pulsars}

\author{Fotis P. Gavriil, Victoria M. Kaspi$^{\dagger}$}
\affil{Department of Physics, Rutherford Physics Building, McGill University, 3600 University Street, Montreal, Quebec, H3A 2T8, Canada }
\author{Deepto Chakrabarty}
\affil{$^{\dagger}$Department of Physics and Center for Space Research, Massachusetts Institute of Technology, Cambridge, MA 02139}

\begin{abstract}
We report on long-term monitoring of anomalous X-ray pulsars (AXPs)
using the \textit{Rossi X-ray Timing Explorer}. 
Using phase-coherent timing, we find a wide variety of 
behaviors among the sources, ranging
from high stability (in \tfn\ and \oft), to instabilities so severe
that phase-coherent timing is not possible (in \tfe).
 We note a
correlation in which timing stability in AXPs decreases with increasing
$\dot{\nu}$.  The timing stability of soft gamma repeaters in
quiescence is consistent with this trend, which is similar to one seen in 
radio pulsars. We consider high
signal-to-noise ratio average pulse profiles 
as a function of energy for each AXP, and find a variety of  behaviors.  
We find no large variability in pulse morphology  nor in  pulsed flux 
as a function of time.
\end{abstract}

\section{Introduction}
Anomalous X-ray pulsars (AXPs) are an unusual class of astrophysical objects. 
There are currently only five confirmed AXPs: \oft,
\tfe, \efo, \soe, and \tfn.  All five are found in the plane of the Galaxy; and two of the five certain members of the class appear
to be located at the geometric centers of apparent supernova remnants.

AXP characteristics can be summarized as follows (see Israel, Mereghetti, \& Stella 2002, for a review): 
they exhibit X-ray pulsations in the
range $\sim$5--12~s; they have pulsed X-ray luminosities in the range
$\sim 10^{34}-10^{35}$~erg~s$^{-1}$; they spin down regularly; their
X-ray luminosities are much greater than the rate of loss of rotational
kinetic energy inferred from the observed spin-down; they have spectra
that are characterized by thermal emission of $kT \sim 0.4$~keV with
evidence for a hard tail in some sources.
Soft gamma repeaters also exhibit AXP-like pulsations in quiescence  
(e.g. Kouveliotou et al. 1998); 
the connection between AXPs and SGRs is intriguing
but not yet clear.  The results here summarize those  reported by Gavriil \& Kaspi (2001).

\section{Results and Discussion}

The results presented here were obtained using the Proportional Counter Array 
(PCA) on board the Rossi X-ray Timing Explorer (\textit{RXTE}). Our observations consist primarily of short snapshots  taken on a monthly basis. In addition, we used a handful of archival observations; the exposures in these observations vary. 

\begin{figure}
\begin{center}
\epsfig{file=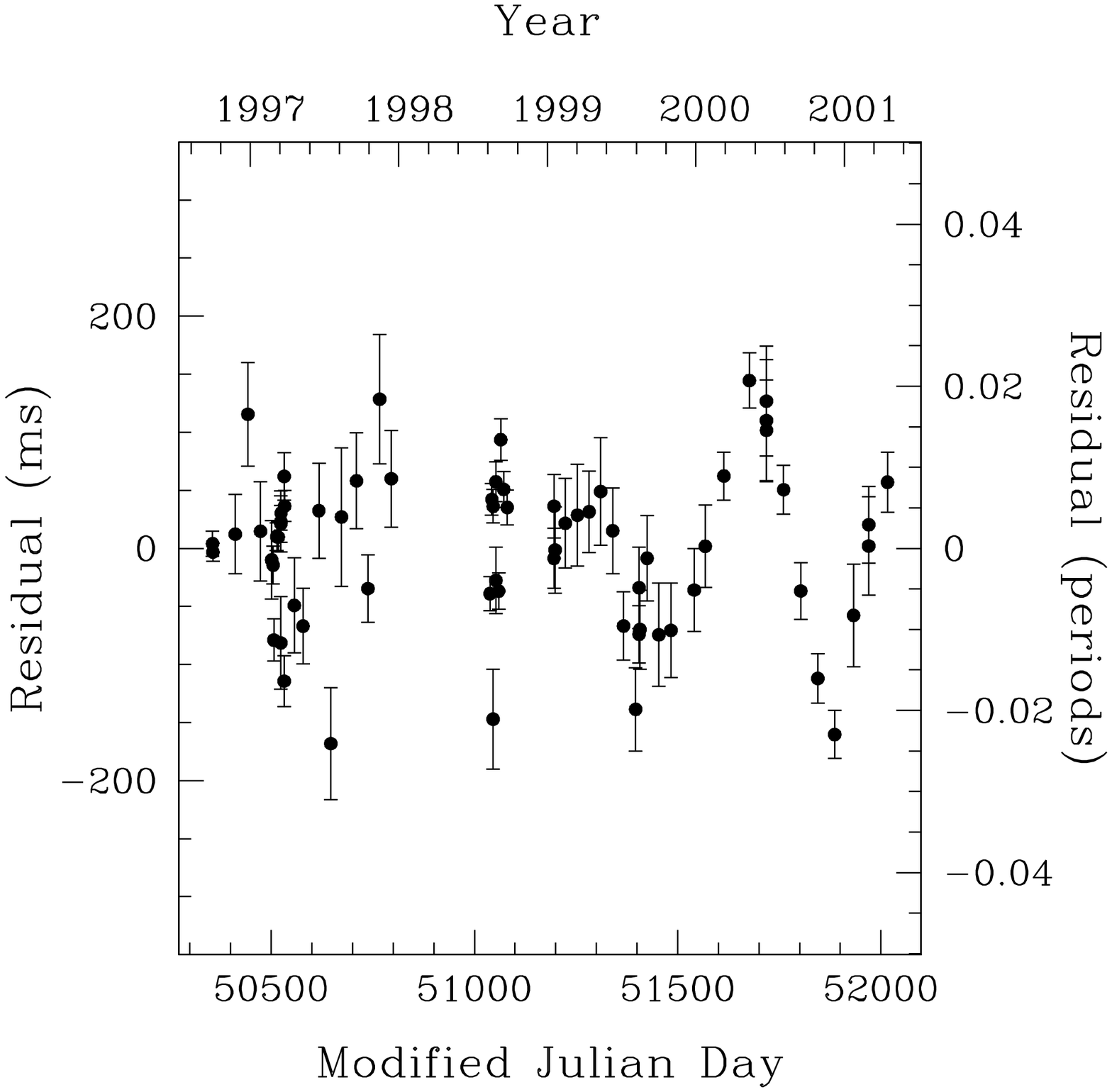,width=1.7in}
\epsfig{file=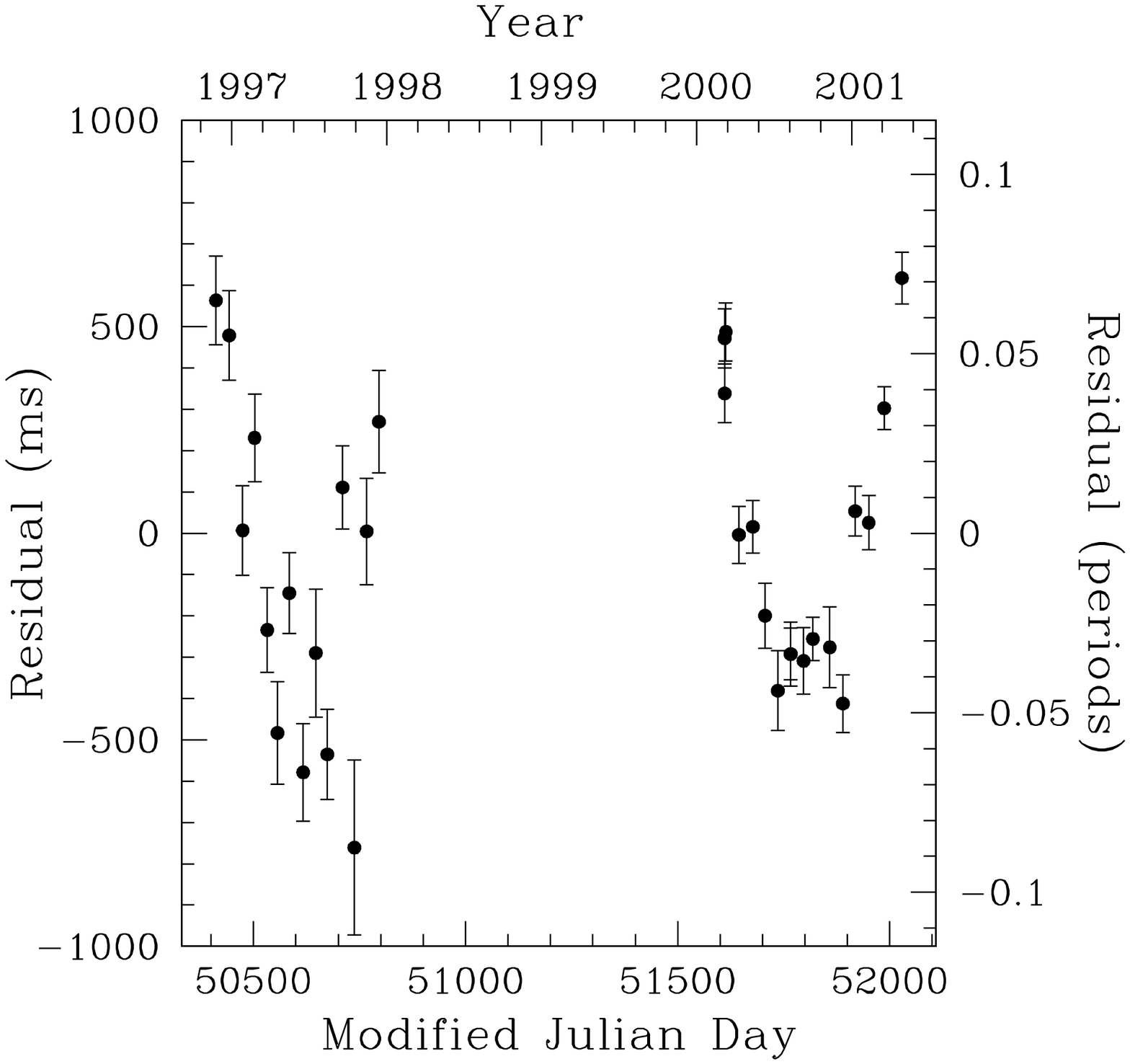,width=1.7in}
\epsfig{file=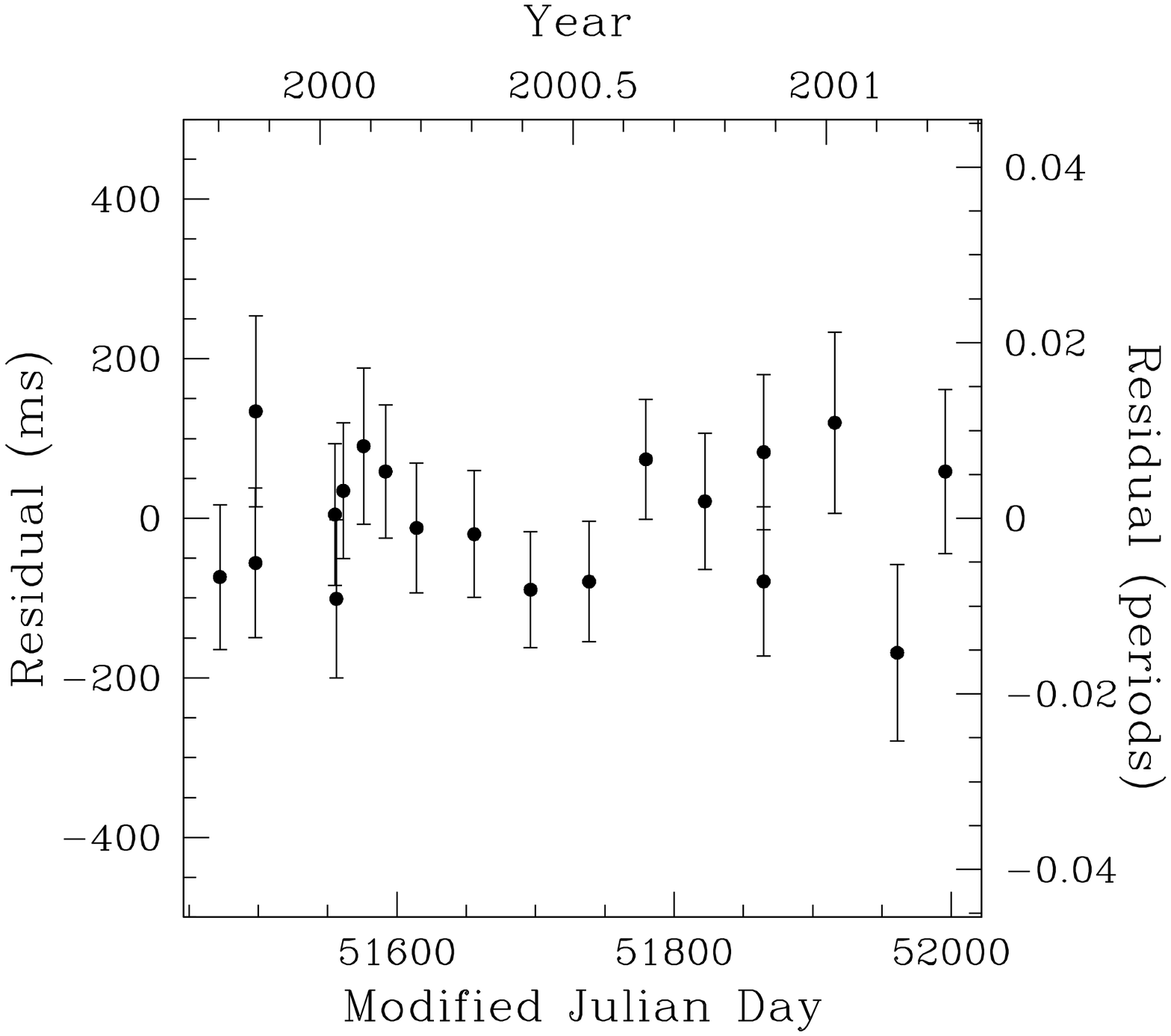,width=1.7in}
\caption{\small Left:  Arrival time residuals for \tfn\ with $\nu$, $\dot{\nu}$ and $\ddot{\nu}$ subtracted; 
Center: Arrival time residuals for \oft\ with $\nu$ and $\dot{\nu}$ subtracted;
Right: Arrival time residuals for \soe\ with $\nu$, $\dot{\nu}$ and $\ddot{\nu}$ subtracted.
\label{fig:res}}
\end{center}
\end{figure}

Phase-coherent timing of AXPs has shown that the rotational stability of some AXPs is comparable to those of some radio pulsars. The rotational stability of \tfn, first reported by Kaspi, Chakrabarty, \& Steinberger (1999), has
now persisted over 4.5~yr, although the inclusion of $\ddot{\nu}$ has recently
been necessary. \oft\ has been an extremely stable
rotator over 4.4~yr of {\it RXTE} monitoring.  For \soe, phase coherent timing has been accomplished in the 1.4~yr
since the glitch reported by Kaspi, Lackey, \& Chakrabarty (2000). In these data, we find a
significant positive $\ddot{\nu}$. This
indicates a decay of the negative $\dot{\nu}$, expected for long-term
glitch recovery, as seen in glitching radio pulsars (e.g. Shemar \& Lyne 1996). \efo\ shows rotational stability but with considerably higher red noise (see Gotthelf et al., this volume). \tfe\ has exhibited far noisier behavior, making  phase-coherent timing using monthly and weekly observations  virtually impossible.  Interestingly there is a correlation between timing stability and $\dot{\nu}$: the sources with the smallest $\dot{\nu}$ are the most stable. Arrival time residuals for \tfn, \oft\ and \soe\ are shown in Figure~\ref{fig:res}.

\begin{figure}
\begin{center}
\epsfig{file=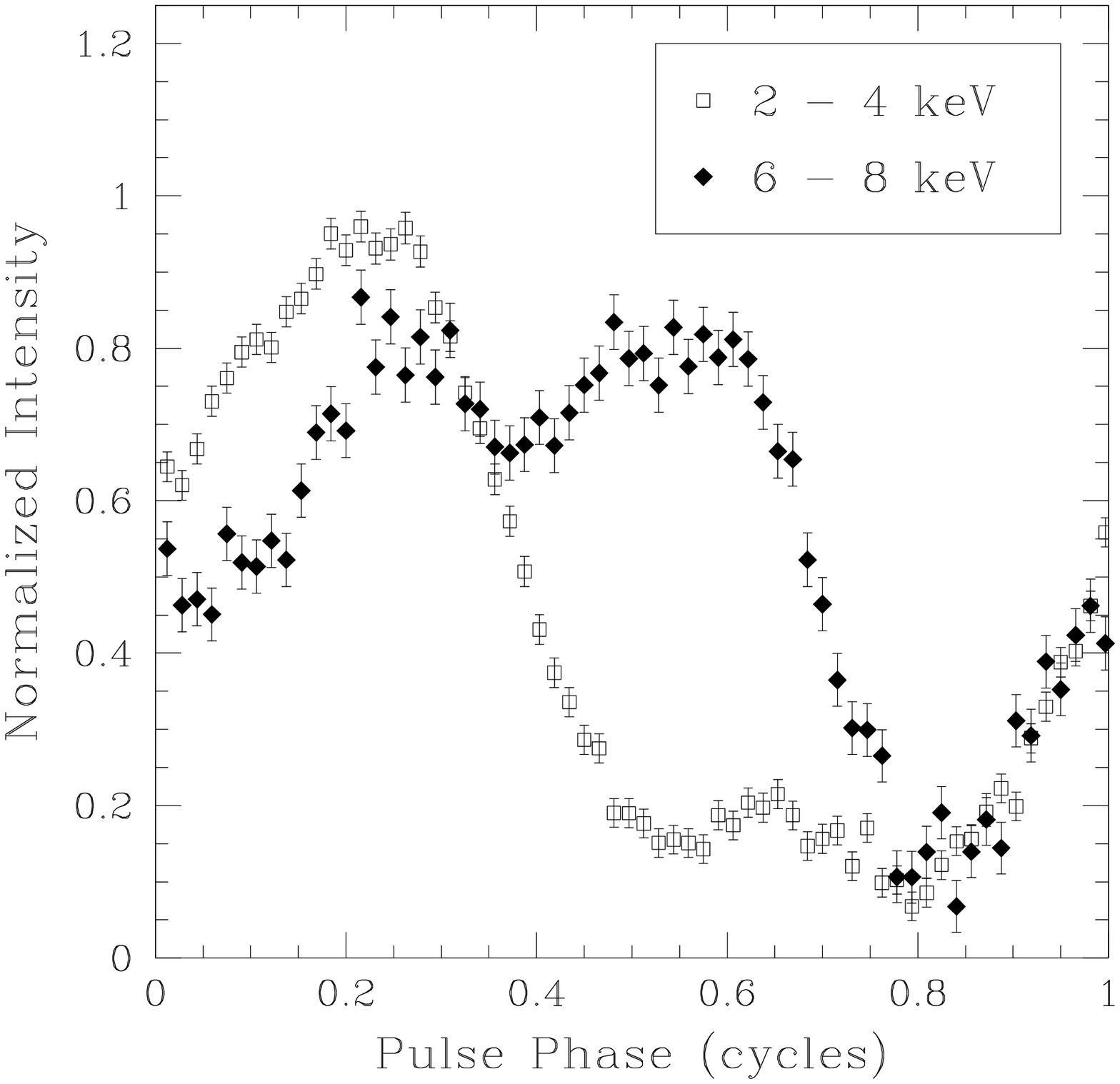,width=1.7in}
\epsfig{file=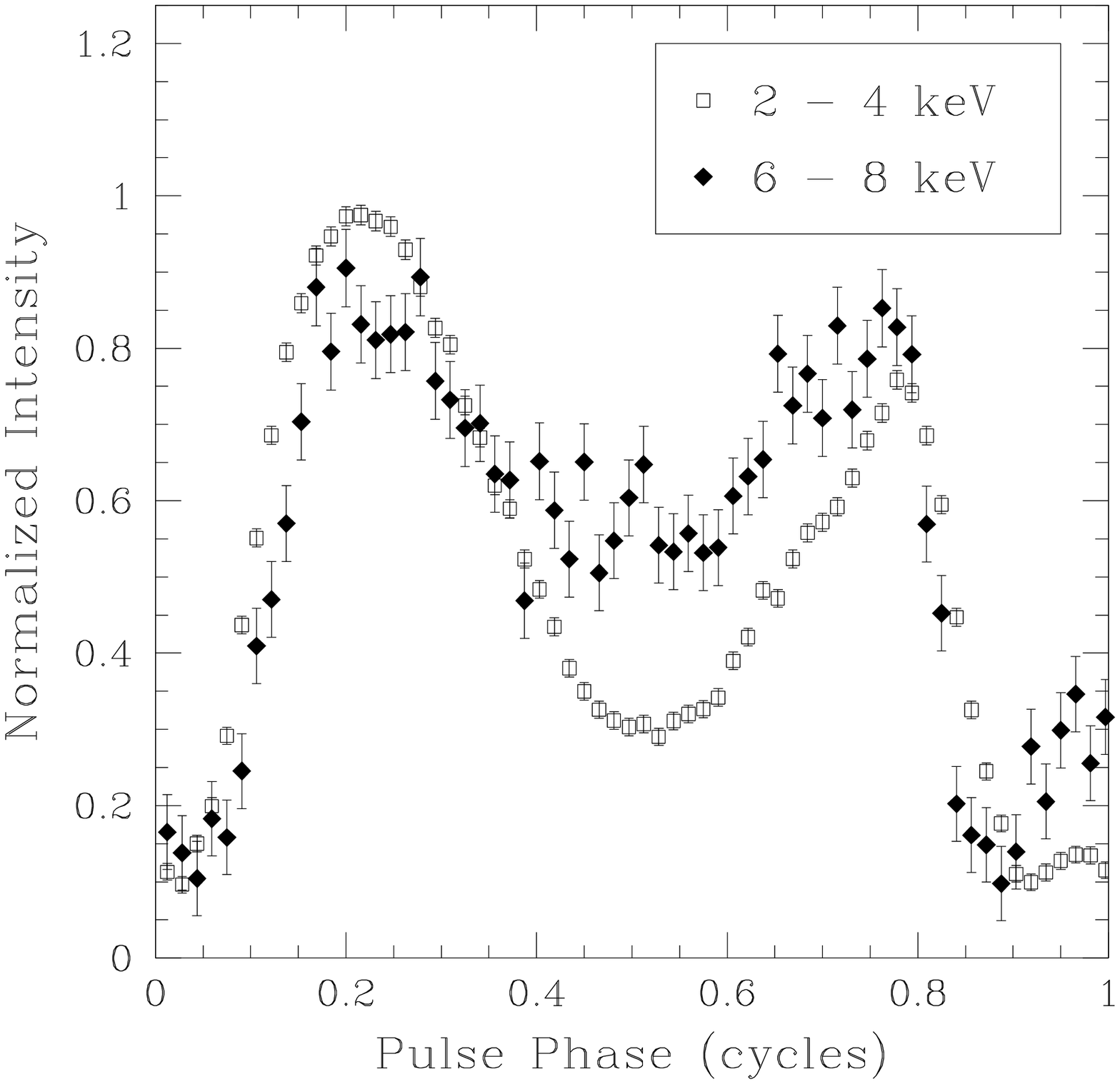,width=1.7in}
\epsfig{file=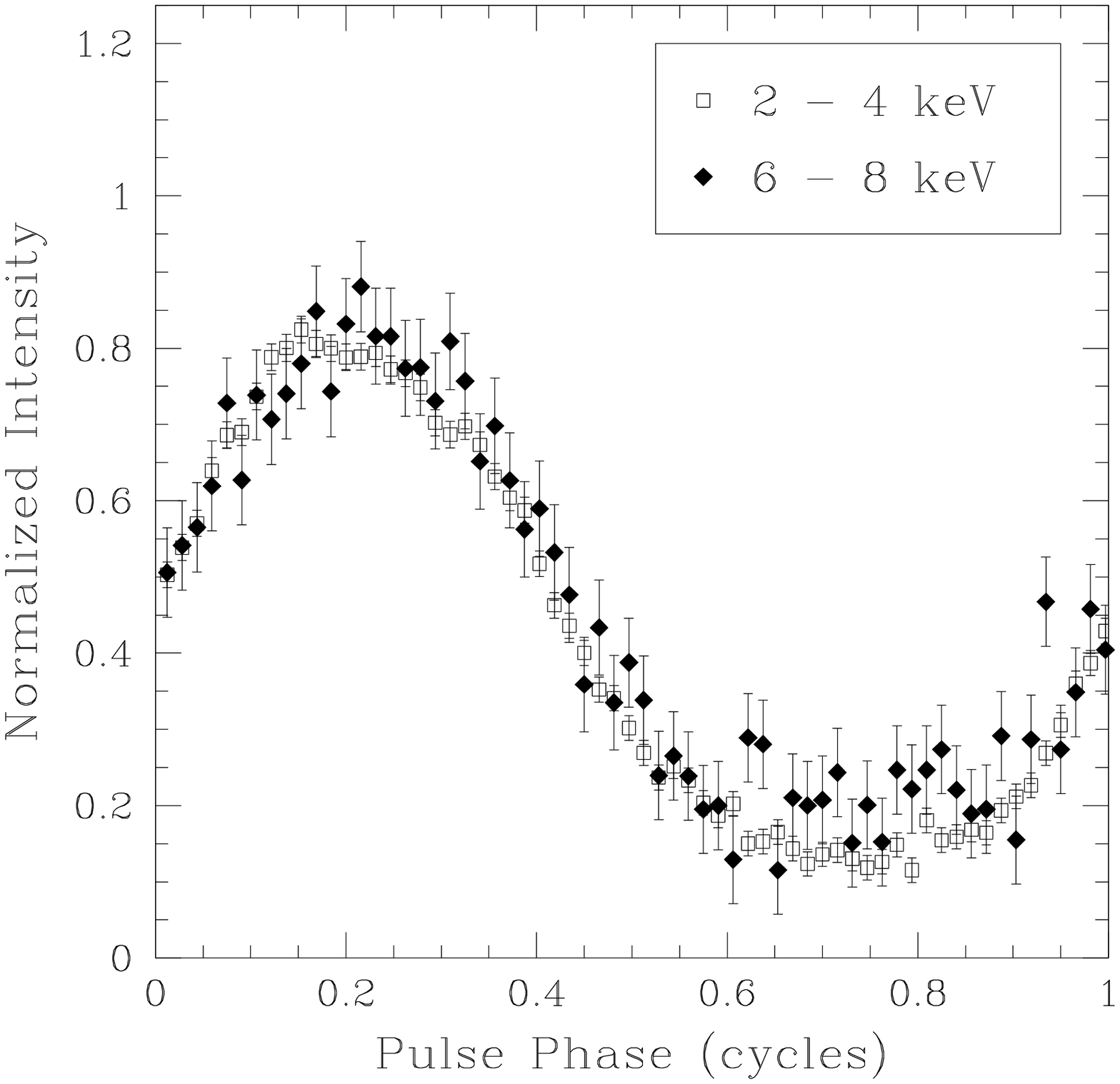,width=1.7in}
\caption{\small Average pulse profile of \soe\ (left), \tfn\ (center) and \tfe\ (right) in two energy bands. Note that the scaling was chosen  to minimize the $\chi^2$ of the  difference between the two profiles.  Thus the only information 
that these plots convey is the relative amplitudes of the features of the profile.
\label{fig:energy}}
\end{center}
\end{figure}

AXP spectra are generally best  fit by a two-component model consisting of a 
photoelectrically absorbed blackbody with a hard power-law tail (Israel et al. 1999). Whether 
these two components are physically distinct is an open question (see \"Ozel, Psaltis, \& Kaspi 2001). To investigate this, we
compared the pulse profile morphology of the AXPs in two energy bands. 
Figure~\ref{fig:energy} displays the average pulse profiles of  
\soe, \tfn\  and \tfe\ in the energy bands 2--4~keV and 6--8~keV.  From Figure~\ref{fig:energy} it is clear that AXP pulse profiles exhibit different degrees of energy dependence;  \soe\ shows high energy dependence, while \tfe\ shows no energy dependence. 

\begin{figure}
\begin{center}
\epsfig{file=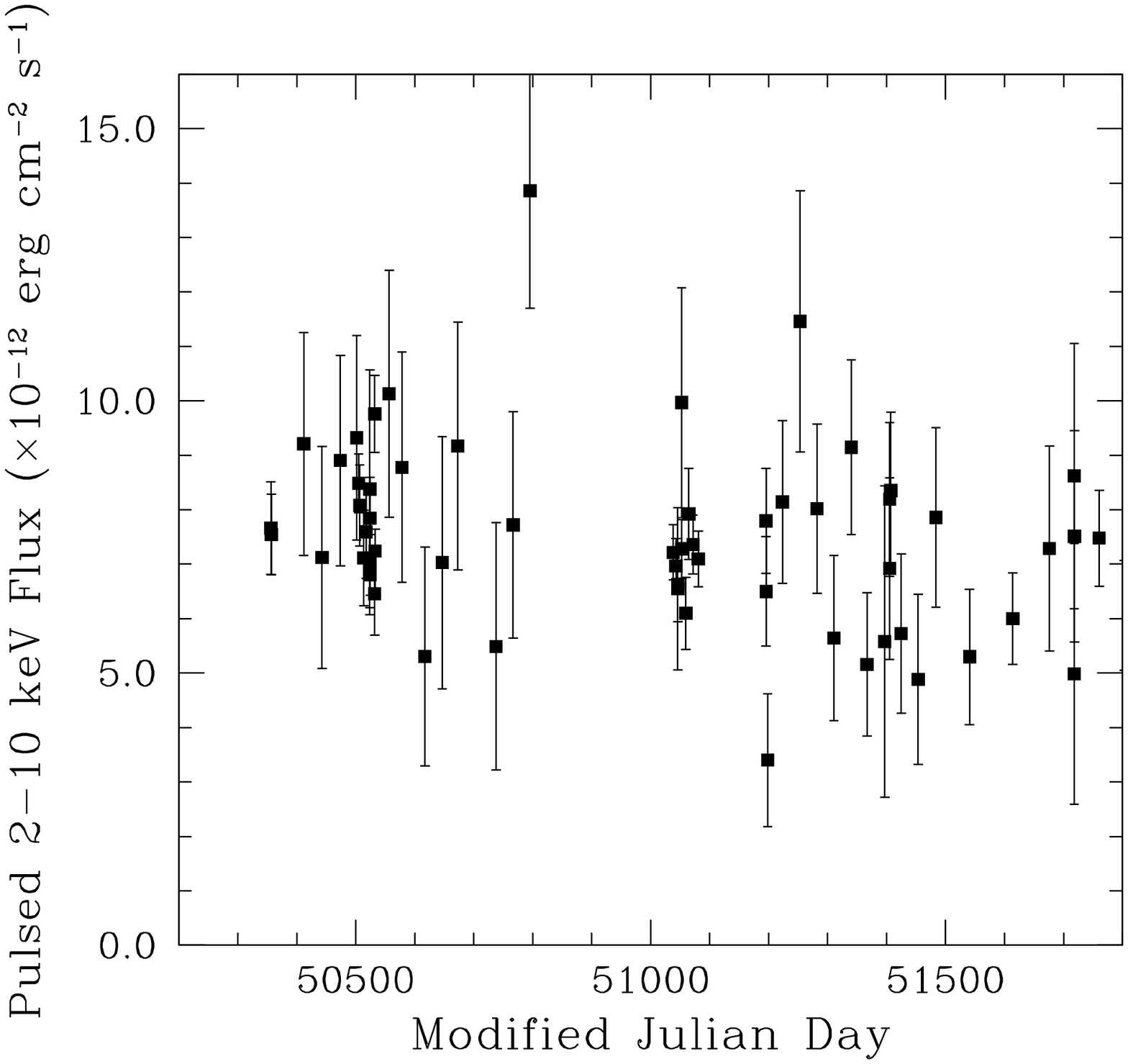,width=1.7in}
\epsfig{file=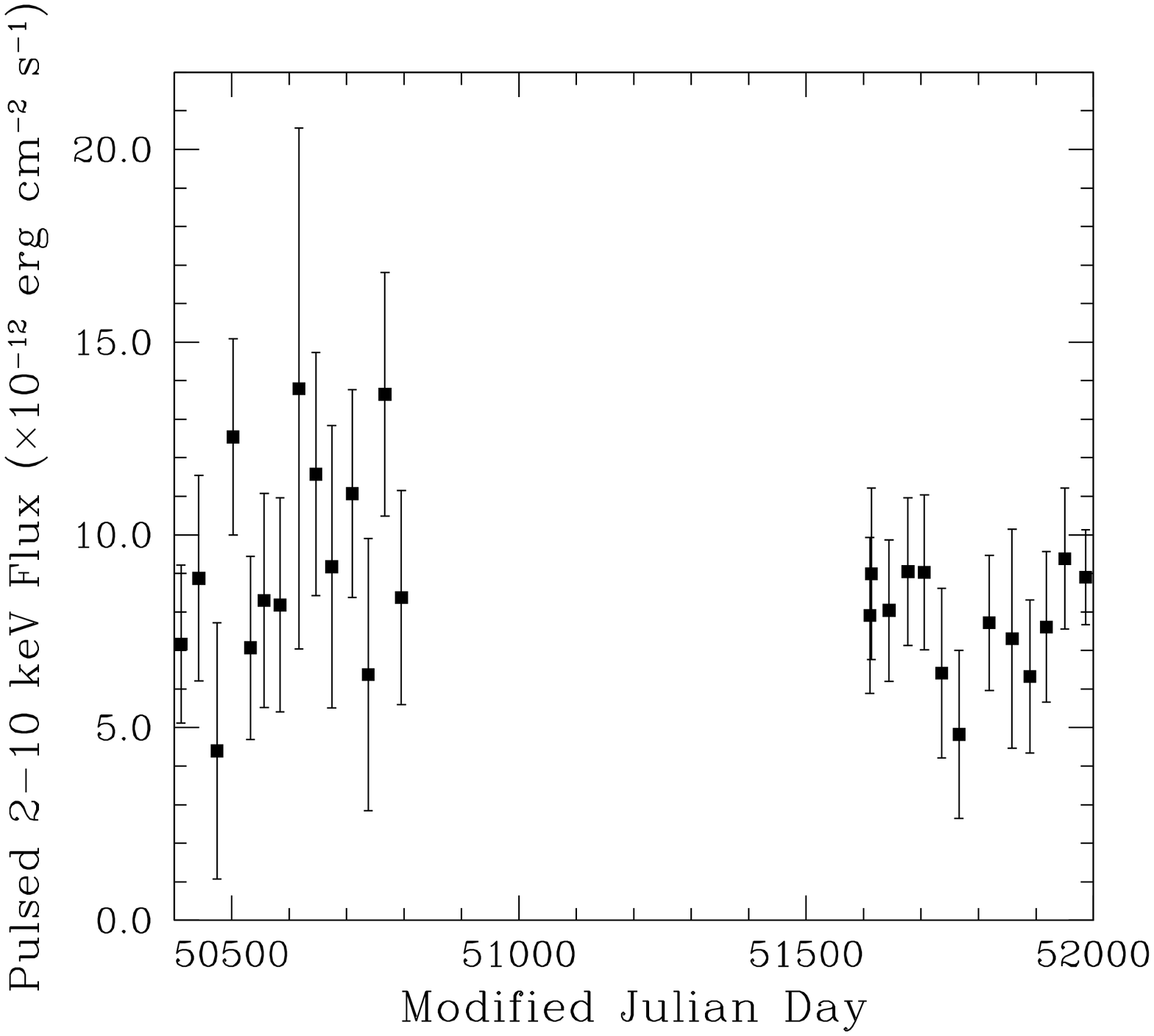,width=1.7in}
\epsfig{file=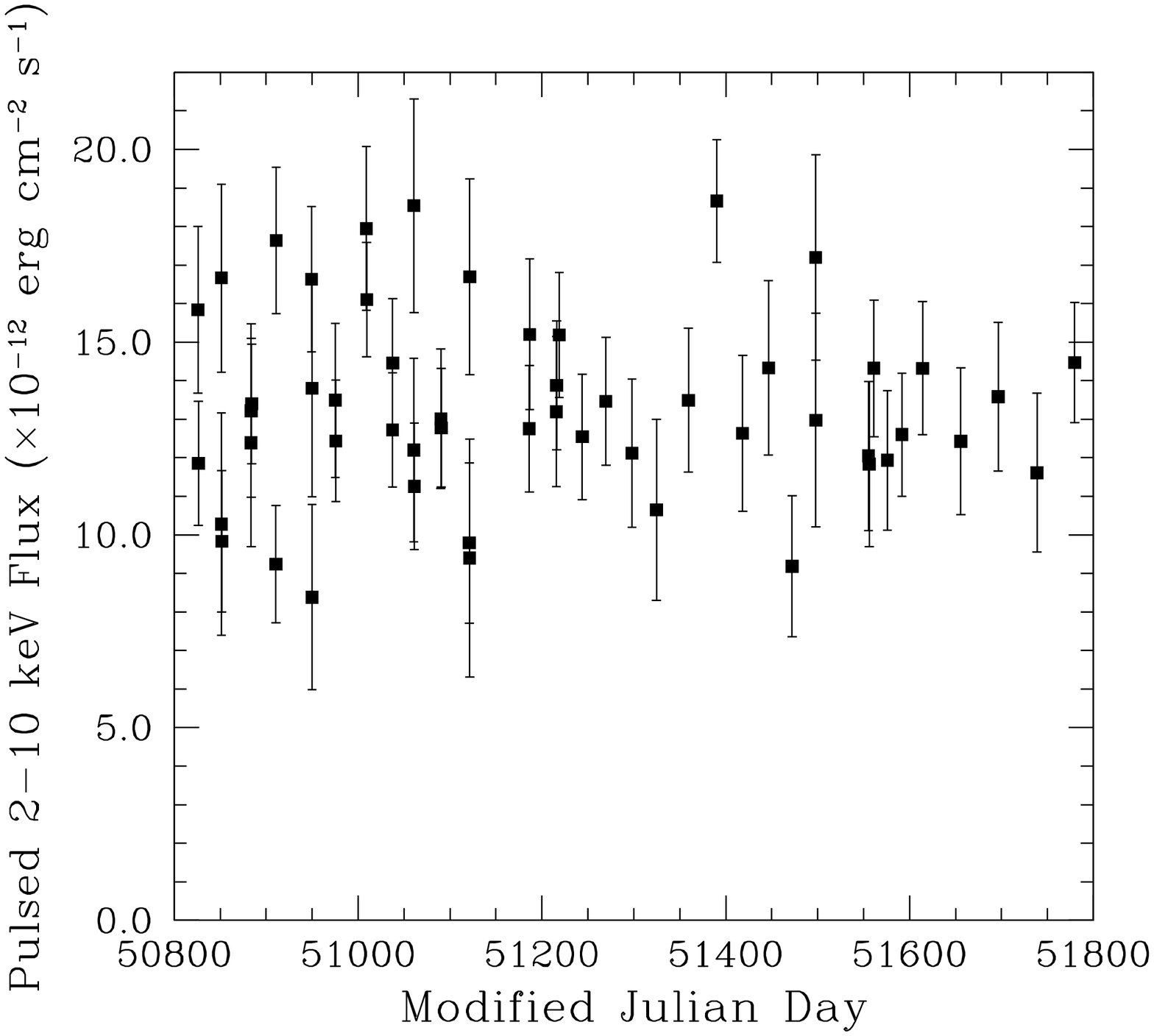,width=1.7in}
\caption{\small Pulsed flux time series for \tfn\ (left), \oft\ (center) and \soe\ (right). Error bars represent 1$\sigma$ confidence intervals.
\label{fig:flux}}
\end{center}
\end{figure}

We have also used our \textit{RXTE} data to monitor the pulsed flux of the AXPs as a function of time. Rapid ($<$ months) flux variability would challenge the magnetar model and accreting sources generally show flux variability correlated with $\dot{\nu}$.
Given the large field-of-view of the PCA and the low count rates for the sources relative to the background, total flux measurements are difficult with our \textit{RXTE} data. Instead, we have determined the pulsed component of the flux, by using the off-pulse emission as a background estimator.
The flux time series for \tfn, \oft\ and \soe\
 are  displayed in Figure~\ref{fig:flux}. 
We do not find evidence for any large variability in the pulsed flux for any source,  and have set $1\sigma$ upper limits on variations
$\sim$20--30\% (depending on the source).  This is surprising given
previous reports of large (factor of 5--10) total flux variations in
\tfn\ and \oft\ (Baykal \& Swank 1996; Oosterbroek et al. 1998).  Assuming a constant pulsed
fraction, this suggests that  more than one of the AXPs happen to be much more
quiescent during the {\it RXTE} monitoring than in the past.

Iwasawa, Koyama, \& Halpern (1992) reported a significant change in the pulse morphology of
\tfn\ in 1.2--14~keV {\it GINGA} observations obtained in 1990, such that the
leading pulse had amplitude roughly half that of the trailing pulse.  
If correct, this has important
implications for the magnetar model, which predicts such pulse
morphology changes in the event of a restructuring of the magnetic
field, as might occur following a major SGR-like outburst. Motivated by this finding, we searched for pulse profile changes in our \textit{RXTE} observations of all the AXPs. We have not detected any large pulse profile
variations. This justifies our other analysis procedures which assume a
fixed profile.  We rule out variations
in features having amplitude $\ga$20\% of the peak amplitude at the
$1\sigma$ level, although the limit depends on source and integration time.

\acknowledgments
 This work was supported in part by a NASA LTSA grant (NAG5-8063) and an NSERC Research Grant (RGPIN228738-00) to VMK, with additional support from a NASA ADP grant (NAG 5-9164). 


\begin{references}
\reference Baykal, A. \& Swank, J. 1996, ApJ, 460, 470
\reference Cordes, J.~M. \& Helfand, D.~J. 1980, ApJ, 239, 640
\reference Gavriil, F.~P. \& Kaspi, V.~M. 2001, ApJ, in press, http://xxx.lanl.gov/abs/astro-ph/0107422
\reference Gotthelf, E.~V., Gavriil, F., Kaspi, V.~M., Vasisht, G., \& Chakrabarty, D. 2001, this volume
\reference Israel, G., Mereghetti, S., \& Stella, L. 2002 (Memorie della Societa' Astronomica Italiana), in press
\reference Israel, G.~L., Oosterbroek, T., Angelini, L., Campana, S., Mereghetti, S., Parmar, A.~N., Segreto, A., Stella, L., Van Paradijs, J., \& White, N.~E. 1999, A\&A, 346, 929
\reference Iwasawa, K., Koyama, K., \& Halpern, J.~P. 1992, PASJ, 44, 9
\reference Kaspi, V.~M., Chakrabarty, D., \& Steinberger, J. 1999, ApJ, 525, L33
\reference Kaspi, V.~M., Lackey, J.~R., \& Chakrabarty, D. 2000, ApJ, 537, L31
\reference Kouveliotou, C., Dieters, S., Strohmayer, T., van Paradijs, J., Fishman, G.~J., Meegan, C.~A., Hurley, K., Kommers, J., Smith, I., Frail, D., \& Murakami, T. 1998, Nature, 393, 235
\reference Oosterbroek, T., Parmar, A.~N., Mereghetti, S., \& Israel, G.~L. 1998, A\&A, 334, 925
\reference  \"Ozel, F., Psaltis, D., \& Kaspi, V.~M. 2001, ApJ, in press, http://xxx.lanl.gov/abs/astro-ph/0105372
\reference Shemar, S.~L. \& Lyne, A.~G. 1996, MNRAS, 282, 677
\end{references}
\end{document}